\documentclass[aps,prl,twocolumn,superscriptaddress,showpacs]{revtex4}
\usepackage{graphicx}
\begin{document}

\title{Etching suspended superconducting tunnel junctions from a multilayer}

\author{H. Q. Nguyen}
\affiliation{Institut N\'eel, CNRS, Universit\'e Joseph Fourier and Grenoble INP, 25 avenue des Martyrs, 38042 Grenoble, France}
\author{L. M. A. Pascal}
\affiliation{Institut N\'eel, CNRS, Universit\'e Joseph Fourier and Grenoble INP, 25 avenue des Martyrs, 38042 Grenoble, France}
\author{Z. H. Peng}
\affiliation{Institut N\'eel, CNRS, Universit\'e Joseph Fourier and Grenoble INP, 25 avenue des Martyrs, 38042 Grenoble, France}
\author{O. Buisson}
\affiliation{Institut N\'eel, CNRS, Universit\'e Joseph Fourier and Grenoble INP, 25 avenue des Martyrs, 38042 Grenoble, France}
\author{B. Gilles}
\affiliation{SIMAP, CNRS, Universit$\acute{e}$ Joseph Fourier and Grenoble INP, 1130 rue de la Piscine, 38402 Saint Martin d'H\`eres, France}
\author{C. B. Winkelmann}
\affiliation{Institut N\'eel, CNRS, Universit\'e Joseph Fourier and Grenoble INP, 25 avenue des Martyrs, 38042 Grenoble, France}
\author{H. Courtois}
\affiliation{Institut N\'eel, CNRS, Universit\'e Joseph Fourier and Grenoble INP, 25 avenue des Martyrs, 38042 Grenoble, France}

\begin{abstract}
A method to fabricate large-area superconducting hybrid tunnel junctions with a suspended central normal metal part is presented. The samples are fabricated by combining photo-lithography and chemical etch of a superconductor - insulator - normal metal multilayer. The process involves few fabrication steps, is reliable and produces extremely high-quality tunnel junctions. Under an appropriate voltage bias, a significant electronic cooling is demonstrated. We analyze semi-quantitatively the thermal behavior of a typical device.
\end{abstract}
\maketitle

Tunnel junctions between a normal metal (N) and a superconductor (S) separated by an insulating oxide barrier (I) are a central component to mesoscopic electronic devices. In SINIS structures with a small metallic island, one can couple superconductivity to single electron effects, which are used for metrological current sources \cite{PekolaNature08}. Such structures have also demonstrated a high potential for on-chip electronic cooling applications \cite{MuhonenReview}. In this case, the superconductor energy gap suppresses tunneling of low-energy electrons, so that only high-energy electrons can tunnel. The normal metal electron population as a whole then reaches a quasi-equilibrium state with a temperature lower than the phonon and the substrate temperatures \cite{NahumAPL94}. Starting from a bath temperature of about 300 mK, an electronic temperature reduction by a factor of about 3 has been achieved in aluminum-copper hybrid devices that are voltage-biased just below the gap of superconducting Al \cite{PekolaPRL04}. Micro-coolers based on large and lithographic junctions have demonstrated a cooling power sufficient for cooling a bulk material \cite{ClarkAPL04} or a detector \cite{MillerAPL08}.

Optimizing a SINIS device for electron cooling can be achieved by increasing the heat current and/or isolating better the cooled electron bath. As the heat current is proportional to the tunnel barriers conductance, reducing the barriers' thickness is the first option. Nevertheless, this can lead to the appearance of two-particle Andreev reflection processes at low energy, which deposit heat in the normal metal \cite{SukumarPRL08}. The obvious alternative is to increase the junction area at a fixed transparency. However, if not epitaxial, large-area solid state tunnel barriers are subject to local fluctuations of the barrier thickness. In this case, most of the current goes through so-called pin-holes \cite{PRL11-Greibe}, giving rise to an unwanted significant sub-gap current that also deposits heat in the normal metal. Depending on the device geometry, the central normal metal island can feature a sizable Ohmic resistance. The large current flowing through a large-area, and thus low-impedance, tunnel junction biased at the gap can then induce a significant Joule dissipation \cite{arxivONeil}. The same current is also expected to heat the superconductor in the vicinity of the junctions area \cite{JAP98-Jochum}. Eventually, for a given cooling power, an efficient electronic cooling relies on a thermal decoupling of cooled electrons and phonons from the environment, including the substrate. It is thus highly desirable to suspend the cooled normal metal over the substrate. 

\begin{figure}[hb]
\begin{center}
\includegraphics[width=0.9\columnwidth,keepaspectratio]{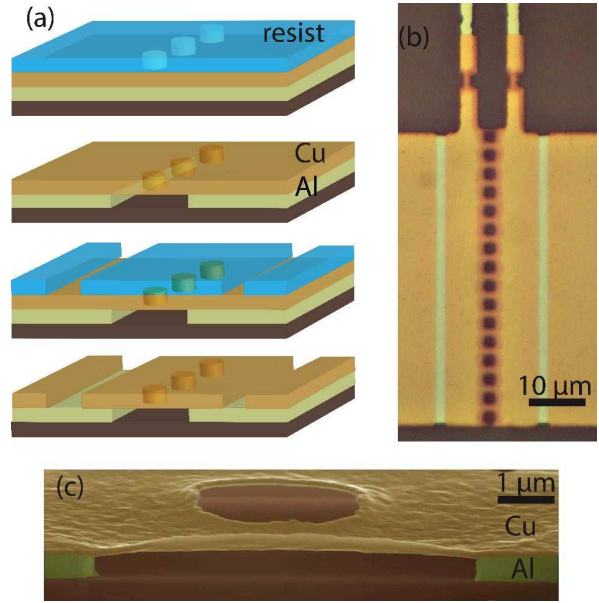}
\caption{(a) From top: fabrication starts with an Al/AlO$_x$/Cu multilayer, on which a photoresist is patterned with contact pads and holes. Then, Cu and Al are successively etched, leaving a suspended membrane of Cu along the line of adjacent holes. A second lithography and etch define the Cu central island. (b) Optical microscope image showing regions by decreasing brightness: bare Al, Cu on Al, suspended Cu and substrate. On the top, two thermometer junctions are added. (c) Colorized scanning electron micrograph of a sample cut using Focused Ion Beam, showing the Cu layer suspended over the holes region. The thickness of Al and Cu is 400 and 100 nm, respectively.}
\label{cap:3d}
\end{center}
\end{figure}

In this Letter, we present a method to fabricate large area SINIS devices of high quality and with a suspended normal metal. The method bases on a pre-deposited multilayer of metals, which can be prepared at the highest quality. The normal part is suspended in the first lithography, which keeps it isolated from the substrate. The second lithography defines the junction area with any geometry of interest. Transport measurements on these junctions show excellent characteristics, without any leakage contribution. Under an appropriate bias, the device demonstrates a significant electronic cooling.

The fabrication starts with depositing a Al/AlO$_x$/Cu multilayer on an oxidized silicon substrate (500 nm SiO$_2$). Prior to deposition, the Si wafer is cleaned by baking it to 300$^o$ C for 4 hours inside an electron beam evaporator, at a pressure below $10^{-9}$ mbar. The wafer is then let to cool down for one hour before depositing Al. The wafer is afterwards moved to a neighboring chamber where it is exposed to a static pressure of oxygen. The later process is known to produce a high-quality thin oxide barrier. The sample is finally moved back to the deposition chamber so that Cu can be deposited, without breaking vacuum.

A first deep ultra-violet lithography \cite{resist} is used to define the overall device geometry. The central part is a series of adjacent holes of diameter 2 $\mu$m and with a side-to-side separation of 2 $\mu$m, see Fig. 1. The copper layer is etched away over the open areas using either Ion Beam Etching (IBE) or wet etching ($\rm{HNO_3}, 65\%$, diluted 1:40 in DI water). Both approaches proved equally satisfactory. The aluminum is then etched through the same resist mask, using a weak base (Microposit MF CD 26 developer, diluted 1:2 in DI water). The etching time (270 s for an Al thickness of 100 nm, or 520 s for 400 nm) is controlled as to completely remove aluminum from the circular region within a horizontal distance of about 2 $\mu$m starting from the hole side. The line of adjacent holes visible in Fig. 1 therefore creates a continuous gap in the Al film, bridged only by a stripe of freely hanging Cu. 
 
The area of the NIS junctions is defined by a second lithography. Through the open areas, trenches are etched into the copper layer only, using one of the two methods cited above. These trenches allow to isolate a copper island, which forms the central normal metal part of the SINIS device. A top view of the complete device is pictured in Fig. 1b. Notably, the suspended copper membrane in the vicinity of the line of holes can be identified by its lower reflectivity with respect to the multilayer. Figure 1c shows a side view scanning electron micrograph of a junction after cutting the sample perpendicular to the line of holes, revealing the vertical structure of the SINIS device, as well as the complete removal of Al below the Cu bridge.

\begin{figure}[htp]
\begin{center}
\includegraphics[width=\columnwidth,keepaspectratio]{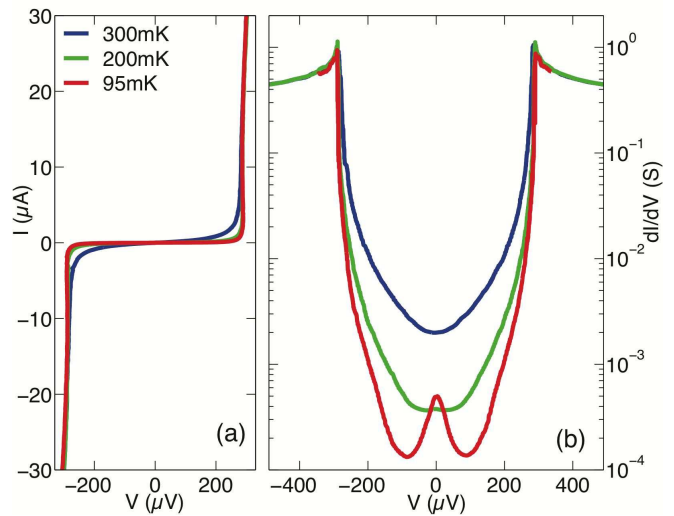}
\caption{(a) Current-voltage characteristic and (b) differential conductance of sample D at different cryostat temperatures, see Table 1 for its fabrication parameters.}
\label{cap:IVs}
\end{center}
\end{figure}

Low temperature transport measurements were performed at a bath temperature down to 100 mK. Accurate electronic filtering of the lines was taken great care of. Twisted pairs of wires were passed through a single CuNi capillary filled with a two-component paste containing radio-frequency absorber \cite{emerson,SlichterAPL09} over 200 cm at the cryostat base temperature. Additionally, two stages of $\pi$-filters were integrated both at room temperature and in the sample holder. Four-point d.c. measurements were performed using home made electronics, in which a current bias is applied to the sample and the voltage is read out.

Fig. \ref{cap:IVs} shows the current-voltage characteristic and the numerically-derived differential conductance of a typical sample at various cryostat temperatures. This plot is actually a combination of multiple curves covering different measurement ranges, extending over several decades of current values. Inside the gap marked by two differential conductance peaks, the sub-gap conductance is found to decrease strongly when temperature is decreased. In a semi-logarithmic representation, an increasing slope of the differential conductance as a function of voltage indicates a decreasing electronic temperature. This behavior is clearly observed, especially close to the gap edge. This demonstrates electronic cooling in the normal metal. 

Similar results were obtained in a series of samples with different fabrication parameters listed in Table I. We have varied the thickness of the Al and Cu layers, as well as oxidation pressure or time in preparing the multilayers. Either dry etching (IBE) or wet etching of Cu was used, without noticeable difference. Electron beam lithography was employed for the second lithography step in samples B and C. All samples show an energy gap $2\Delta\sim$ 350 $\mu$eV, consistent with the bulk gap of aluminum. The junctions' normal state resistances $R_N$ roughly scale with the inverse junction area and increase with oxidation pressure.

\begin{table}[b]
\begin{tabular}{cccccccc}\hline
	id & $t_{Al}$&$t_{Cu}$&$P_{O_2}$ & Area & $R_N$ & $2\Delta$ & $G_{min}/G_N$\\
	 & (nm) & (nm) & (mbar) & $(\mu m^2)$ & $(\Omega)$ & $(\mu eV)$ &  \\
	\hline\hline
	A & 100 & 100 & 4000 & 1200 & 14.6 & 355 & 2.2x10$^{-5}$ \\
	B & 100 & 50 & 5 & 100 & 64 & 356  & 5x10$^{-5}$ \\
	C & 100 & 50 & 100 & 200 & 31.8 & 337 & 2.1x10$^{-4}$\\
	D & 100 & 50 & 1 & 200 & 2.8 & 360 & 3.7x10$^{-4}$\\
	E & 400 & 100 & 100 & 400 & 33 & 346 & 5.5x10$^{-5}$\\
	F & 400&100& 1 & 200 & 7.8 & 321 & 2x10$^{-4}$\\
	G & 400 & 100 & 1 & 650 & 1.7 & 338 & 8x10$^{-4}$ \\\hline
\end{tabular}
\caption{Parameters of different SINIS devices: $t_{Al}$ and $t_{Cu}$ are the thickness of the Al and Cu layers, respectively. $P_{O_2}$ is the oxidation pressure. All samples were oxidized for 15 minutes, except for sample A that was oxidized for 30 minutes. $R_N$ is the normal-state resistance. The values of the sub-gap conductance ratio $G_{min}/G_N$ were extracted from measurements below 100 mK (A to E) or 150 mK (F, G).}
\end{table}

At very low temperature, a zero-bias peak of the differential conductance is observed (see Fig. \ref{cap:IVs}), which is a clear signature of coherent Andreev processes \cite{SukumarPRL08}. It defines a higher bound for the ratio $G_{min}/G_N$ of the minimum differential conductance $G_{min}$ to the high-bias conductance $G_N$, thereby characterizing the contribution of possible pin-holes. Based on our measured $G_{min}/G_N$ values in the $10^{-4}$ to $10^{-5}$ range, we claim that our junctions, although not epitaxial, are pinhole-free. Furthermore, while we were initially concerned that redeposition of the etched metals might shunt the junction, this was not observed. Samples do not show any sign of aging, even when stored in ambient conditions for months.

Let us now discuss semi-quantitatively Fig. \ref{cap:IVs} data in terms of electronic thermal behavior. We have measured the electronic temperature as a function of the cooler voltage bias by using the two attached smaller junctions, see Fig. \ref{cap:3d}b, as an electron thermometer (data not shown). The electronic temperature was determined by fitting the measured differential conductance characteristics as a function of the voltage with the theoretical expectation, i.e. a thermally-smeared BCS density of states. At a bath temperature $T_{bath}$ of 300 mK and at the optimum bias point, the measured electronic temperature $T_e$ goes down to 240 mK. 

In the hypothesis that phonons in the normal metal remain at the bath temperature, the related electron-phonon coupling power $P_{e-ph}=\Sigma U(T_e^5-T_{bath}^5)$ \cite{MuhonenReview} amounts to 33 pW. Here $U$ is the normal metal island volume and we take the established value $\Sigma=$2 nW$\mu m^{-3}K^{-5}$ for the electron-phonon coupling parameter in Cu. The Andreev heat also brings a sizable contribution, with a power estimated as $G_{min}\Delta^2/e^2$ of about 20 pW. In comparison, the Joule heat dissipated in the normal metal island resistance, whose value is about 0.02 $\Omega$, by the bias current of 10 $\mu A$ at the optimum point is of the order of 2 pW, which is almost negligible. The total estimated heat load thus amounts to about 55 pW.

The cooling power $P_{cool}$ can be theoretically calculated from the expression \cite{PhysicaUllom}:
\begin{equation}
P_{cool}=\frac{\Delta^2}{e^2R_N}\{0.59(\frac{k_BT_{e}}{\Delta})^{3/2}-\sqrt{\frac{2 \pi k_BT_{s}}{\Delta}}\exp{-\frac{\Delta}{k_BT_{s}}}\},
\end{equation}
where $T_s$ is the superconductor temperature. Let us first consider the equilibrium hypothesis where the superconductor is well thermalized at the bath temperature. At a bath temperature of 0.3 K, the second term in Eq. (1) is then almost negligible. Taking an electronic temperature $T_e$ of 0.24 K, we obtain a cooling power of about 520 pW per junction, \mbox{i.e.} 1040 pW in total. The discrepancy with the estimation of the total heat load is presumably due to the overheating of the superconducting electrodes submitted to a large quasi-particle current \cite{arxivONeil,JAP98-Jochum,PRB12-Rajauria}. This effect actually shows up experimentally as a slight back-bending feature on the current-voltage characteristic close to the gap edge. The related superconductor temperature $T_s$ can be estimated by using it as a free parameter in Eq. (1) so that the calculated cooling power fits the estimated heat load. In this case, the second term is obviously significant. In this way, we obtain $T_s$ = 0.58 K. This value appears as reasonable and could be compared to theoretical predictions similar to the one developed in Ref. \cite{PRB12-Rajauria}, but using a 2D formalism.

Removing the contact between the substrate and the cooled metal by suspending the latter is quite promising for electronic refrigeration applications as it can significantly improve cooling of electrons and phonons. Such suspended metallic beams have been made using usual shadow evaporation technique and etch of the underneath layer using high pressure Reactive Ion Etching either on bulk substrates or membranes \cite{APL-Paraonua,LiAPL07,MuhonenAPL09,KoppinenPRL09}. Every approach starts with patterning a resist bilayer, which leaves little room for substrate cleaning before deposition. The resist itself can also pollute the metal structures deposited. Moreover, the junction dimensions cannot usually be pushed well beyond the resist thickness. Eventually, the RIE etch dictates some material choices.

In comparison, our approach has several advantages. As fabrication starts with preparing the multilayer, the wafer can be baked in ultra-high vacuum environment, which we believe to be an essential ingredient for obtaining pinhole-free NIS junctions. Deposition can be made at high temperature, which enables epitaxial growth of Al and a high oxide quality. The layers' thicknesses and the junction dimension can be increased independently of any resist thickness. Other material combinations than Cu and Al may prove interesting, provided an appropriate method of selective etching and over-etching of the two different materials exists. 

In summary, we have presented a method for fabricating high-quality and large-area SINIS junctions, combining simple photo-lithography and etching techniques. The process yields a suspended normal metal membrane, bridging superconducting leads. The junctions are of high quality with no pin-holes or unwanted shunt. When biased with a voltage just below the superconductor gap, our samples display a significant electronic cooling. A preliminary thermal analysis suggests that overheating in the superconducting electrodes is responsible for the moderate cooling amplitude. Further investigation is currently carried on to optimize heat dissipation near the SINIS junctions. 

Our work was funded by the EU FRP7 low temperature infrastructure MICROKELVIN and by the SOLID project. Samples have been fabricated at Nanofab facility at Institut N\'eel. The authors thank B. Pannetier, T. Fournier, T. Crozes, J.-F. Motte, S. Dufresnes, J. Muhonen, M. Meschke and J. P. Pekola for help and discussions.

\end{document}